\documentclass[prb,twocolumn,showpacs,preprintnumbers,amsmath,amssymb]{revtex4}

\usepackage{graphicx}
\usepackage{dcolumn}
\usepackage{bm}

\begin{document}

\title{Effect of doping and disorder on the half-metallicity of full Heusler alloys}

\author{I. Galanakis$^1$, K. \"Ozdo\~gan$^2$, E. \c Sa\c s\i o\~glu$^{3,4}$,
and B. Akta\c s$^2$}
\email{galanakis@upatras.gr,kozdogan@gyte.edu.tr,ersoy@mpi-halle.de}

\affiliation{ $^1$ Department of Materials Science, School of
Natural  Sciences, University of Patras,  GR-26504 Patra, Greece\\
$^2$ Department of Physics, Gebze Institute of Technology, Gebze,
41400, Kocaeli, Turkey \\ $^3$ Max-Planck Institut f\"ur
Mikrostrukturphysik, D-06120 Halle, Germany\\
$^4$ Fatih University,  Physics Department, 34500,    B\" uy\"
uk\c cekmece,  \.{I}stanbul, Turkey}

\date{\today}

\begin{abstract}
Heusler alloys containing Co and Mn are amongst the most heavily
studied half-metallic ferromagnets for future applications in
spintronics. Using state-of-the-art electronic structure
calculations, we investigate the effect of doping and disorder on
their electronic and magnetic properties. Small degrees of doping
by substituting Fe or Cr for Mn scarcely affect the
half-metallicity. A similar effect is also achieved by mixing the
sublattices occupied by the Mn and sp atoms. Thus the
half-metallicity is a robust property of these alloys.
\end{abstract}


\maketitle

The intensive development of electronics based on the combination
of magnetic and semiconducting materials has brought in the center
of scientific research new exotic materials. Half-metallic
ferromagnets, which were first predicted by de Groot and
collaborators in 1983,\cite{deGroot} have the peculiarity that the
band-structure of the minority-spin electrons is semiconducting
while of the majority-spin electrons is a normal metallic one.
Such materials could maximize the efficiency of spintronic
devices.\cite{Zutic} Several Heusler compounds like NiMnSb and
Co$_2$MnSi have been  predicted to be half-metals.\cite{Galanakis}

Ishida and collaborators were, to the best of our knowledge, the
first to study by means of \textit{ab-initio} calculations the
full-Heusler compounds of the type Co$_2$MnZ, where Z stands for
Si and Ge, and have shown that they are half-metals.\cite{Ishida}
Later the origin of half-metallicity in these compounds has been
largely explained.\cite{Galanakis} Many experimental groups during
the last years have worked on these compounds and have tried to
synthesize them mainly in the form of thin films and incorporate
them in spintronic devices.  The group of Westerholt  has
extensively studied the properties of Co$_2$MnGe films and they
have incorporated this alloy in the case of spin-valves and
multilayer structures.\cite{Westerholt} The group of Reiss managed
to create magnetic tunnel junctions based on
Co$_2$MnSi.\cite{Reiss} A similar study of Sakuraba and
collaborators resulted in the fabrication of magnetic tunnel
junctions using Co$_2$MnSi as one magnetic electrode and Al-O as
the barrier (Co$_{75}$Fe$_{25}$ is the other magnetic electrode)
and their results are consistent with the presence of
half-metallicity for Co$_2$MnSi.\cite{Sakuraba} Dong and
collaborators recently managed to inject spin-polarized current
from Co$_2$MnGe into a semiconducting structure.\cite{Dong}
Finally Kallmayer \textit{et al.} studied the effect of
substituting Fe for Mn in Co$_2$MnSi films and have shown that the
experimental extracted magnetic spin moments are compatible with
the half-metallicity for small degrees of doping.\cite{Kallmayer}

It is obvious from the experimental results that the full-Heusler
compounds containing Co and Mn are of particular interest for
spintronics. Not only they combine high Curie temperatures and
coherent growth on top of semiconductors (they consist of four fcc
sublattice with each one occupied by a single chemical element)
but in real experimental situations they can preserve a high
degree of spin-polarization at the Fermi level. In order to
accurately control their properties it is imperative to
investigate the effect of defects, doping and disorder on their
properties. Recently Picozzi \textit{et al.} published a study on
the effect of defects in Co$_2$MnSi and Co$_2$MnGe.\cite{Picozzi}
Our work aims to further study the effect of doping and disorder
on the electronic and magnetic properties of such compounds.
Doping is simulated by substituting Fe or Cr for Mn while disorder
occurs between the Mn and the sp atom. The electronic structure
calculations are performed using the full--potential nonorthogonal
local--orbital minimum--basis band structure scheme
(FPLO).\cite{koepernik} Details of similar type of calculations
have been published elsewhere.\cite{Ozdogan}

\begin{figure}
\includegraphics[scale=0.5]{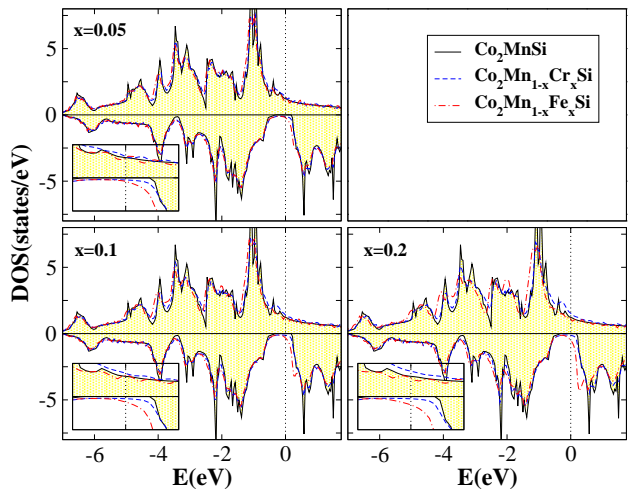}
\caption{(Color online)Spin-resolved total density of states (DOS)
for the case of Co$_2$Mn$_{1-x}$Cr$_x$Si and
Co$_2$Mn$_{1-x}$Fe$_x$Si for three difference values of the doping
concentration $x$. DOS's are compared to the one of the undoped
Co$_2$MnSi alloy. In the onsets we have blown up the region around
the Fermi level (which we have set as the zero of the Energy
axis). Note that positive values of DOS refer to the majority-spin
electrons and negative values to the minority-spin electrons.}
\label{fig1}
\end{figure}

\begin{table}
\caption{Total and atom-resolved spin magnetic moments for the case
of Fe and Cr doping of the Mn site in $\mu_B$. The total moment in
the cell is the sum of the atomic ones multiplied by the
concentration of this chemical element.} \label{table1}
\begin{ruledtabular}
 \begin{tabular}{l|ccccc|ccccc}
    &    Total  &   Co &   Mn&Cr&
 sp &  Total &
Co &   Mn&Fe&  sp \\ \hline
$x$& \multicolumn{5}{c|}{Co$_2$Mn$_{1-x}$Cr$_{x}$Si} & \multicolumn{5}{c}{Co$_2$Mn$_{1-x}$Fe$_{x}$Si}  \\
   0.00 &   5.00 &   1.96 &    3.13 &        &      -0.09 &   5.00  & 1.96&   3.13   &       &
  -0.09\\
  0.05  &  4.95 &   1.97  &   3.12 &    2.06 &    -0.09 &   5.05   & 2.02   &  3.13    &   2.87  &
  -0.09\\
  0.10  &  4.90&   1.97   &  3.12  &   2.07   &  -0.09&   5.09   & 2.06    & 3.17  &   2.85  &
  -0.08\\
  0.20  &  4.80&    1.97  &   3.12 &    2.09    &  -0.08&   5.14   & 2.13 &  3.16  &   2.82   &
  -0.08\\
 $x$&\multicolumn{5}{c|}{Co$_2$Mn$_{1-x}$Cr$_{x}$Ge} &\multicolumn{5}{c}{Co$_2$Mn$_{1-x}$Fe$_{x}$Ge}   \\
  0.00   & 5.00   &1.87  & 3.20    &         & -0.06 &  5.00 &  1.87  &  3.20   &        &    -0.06\\
  0.05   & 4.95  & 1.86  & 3.21 & 2.05   & -0.06&  5.05  &  1.91  &  3.22 &   2.88 &
  -0.06\\
  0.10   & 4.90 &  1.86  & 3.22    & 2.07   & -0.06&  5.10 &  1.96  &  3.23  &   2.88  &   -0.06\\
  0.20    &4.80&   1.86  & 3.22   & 2.10   & -0.06&  5.19  &  2.06  &  3.26  &  2.89   &   -0.05\\
 $x$&\multicolumn{5}{c|}{Co$_2$Mn$_{1-x}$Cr$_{x}$Sn} & \multicolumn{5}{c}{Co$_2$Mn$_{1-x}$Fe$_{x}$Sn}  \\
  0.00    &5.02   &   1.78&    3.32  &      &    -0.08& 5.02  &  1.78    & 3.32      &       &
  -0.08\\
  0.05    &4.98  &    1.77 &  3.34  &  2.24  &  -0.08 & 5.06   & 1.82   & 3.35   &  2.89 &  -0.08\\
  0.10    &4.92 &    1.77  &   3.34 &    2.24  &  -0.08& 5.11  &  1.87   & 3.36    &   2.90 &  -0.07\\
  0.20    &4.82&     1.76  &  3.35 &    2.27    & -0.08& 5.20 &   1.98   & 3.38 &    2.91  &
  -0.07
\end{tabular}
\end{ruledtabular}
\end{table}

The first part of our investigation concerns the doping of
Co$_2$MnSi, Co$_2$MnGe and Co$_2$MnSn. To simulate the doping by
electrons we substitute Fe for Mn while to simulate the doping of
the alloys with holes we substitute Cr for Mn. We study the cases
of moderate doping substituting 5\%, 10\% and 20\% of the Mn
atoms. The use of coherent potential approximation in our
calculations ensures that the doping is performed in a random way.
In Table \ref{table1} we have gathered the total and atom-resolved
spin moments for all cases under study and in Fig. \ref{fig1} the
total density of states (DOS) for the Co$_2$Mn$_{1-x}$Fe$_x$Si and
Co$_2$Mn$_{1-x}$Cr$_x$Si compounds blowing up in the onsets the
region around the Fermi level where the gap exists.

We will start our discussion from the DOS presented in Fig.
\ref{fig1}. As discussed in Ref. \onlinecite{Galanakis} the gap is
created between states located exclusively at the Co sites. The
states low in energy (around -6 eV) originate from the low-lying
$p$-states of the $sp$ atoms (there is also an $s$-type state very
low in energy which is not shown in the figure). The majority-spin
occupied states form a common Mn-Co band while the occupied
minority states are mainly located at the Co sites and minority
unoccupied at the Mn sites. Doping the perfect ordered alloy with
either Fe or Cr first smoothens the valleys and picks along the
energy axis. This is a clear sign of the chemical disorder; Fe and
Cr induce picks at slightly different places than the Mn atoms
resulting to this smoothening and as the doping increases this
phenomenon becomes more intense. The important detail is what
happens around the Fermi level and in what extent is the gap in
the minority band affected by the doping. So now we will
concentrate only at the enlarged regions around the Fermi level.
The blue dashed lines represent the Cr-doping while the red
dash-dotted lines are the Fe-doped alloys. Cr-doping has only
marginal effects to the gap. Its width is narrower with respect to
the perfect compounds but overall the compounds retain their
half-metallicity. In the case of Fe-doping the situation is more
complex. Adding electrons to the system means that, in order to
retain the perfect half-metallicity, these electrons should occupy
high-energy lying antibonding majority states. This is
energetically not very favorable and for these moderate degrees of
doping a new shoulder appears in the unoccupied states which is
close to the right-edge of the gap; a sign of a large change in
the competition between the exchange splitting of the Mn majority
and minority states and of the Coulomb repulsion. In the case of
the 20\% Fe doping this new peak crosses the Fermi level and the
Fermi level is no more exactly in the gap but slightly above it.
Further substitution should lead to the complete destruction of
the half-metallicity as in the Quaternary Heusler alloys with a
Mn-Fe disordered site.\cite{GalaQuat}

\begin{figure}
\includegraphics[scale=0.5]{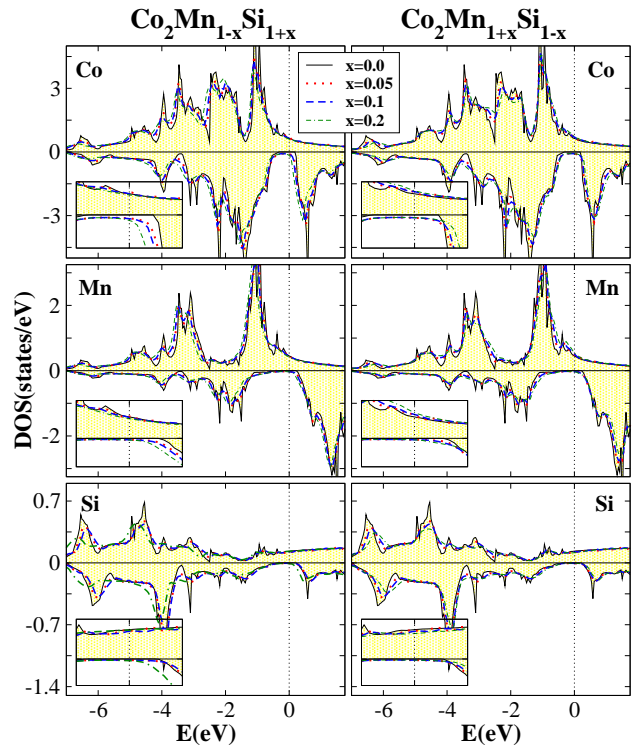}
\caption{(Color online) Atom resolved DOS for the cases of Si
(left panel) and Mn (right panel) excess in Co$_2$MnSi alloy
with respect to the perfect one ($x$=0). In the onsets we have
blown up the region around the Fermi level. \label{fig2}}
\end{figure}

\begin{table*}
\caption{Total and atom-resolved spin magnetic moments for the case
of Mn-sp atom disorder in $\mu_B$. The total moment in the cell is
the sum of the atomic ones multiplied by the concentration of this
chemical element.} \label{table2}
\begin{ruledtabular}
 \begin{tabular}{r|cccc|cccc|cccc}
 & \multicolumn{4}{c|}{Co$_2$Mn$_{1+x}$Si$_{1-x}$}
 & \multicolumn{4}{c|}{Co$_2$Mn$_{1+x}$Ge$_{1-x}$} &
 \multicolumn{4}{c}{Co$_2$Mn$_{1+x}$Sn$_{1-x}$}\\ \hline
  $x$ &    Total  &   Co &   Mn&  Si &    Total  &   Co &   Mn&  Ge&    Total  &   Co &   Mn&
  Sn\\ \hline
 -0.20   & 4.40&   1.92 &  3.19  &  -0.06&    4.40  &   1.83&   3.29  &     -0.05&   4.41 &   1.78&    3.41  &   -0.08\\
 -0.10   & 4.70&   1.95    & 3.15 &    -0.08&  4.70 &1.84  &  3.25  &   -0.06&4.73 &  1.75&    3.41   &  -0.08\\
 -0.05   & 4.85&   1.96  & 3.14  &  -0.08&  4.85  &   1.85 &  3.23   & -0.06& 4.88  &  1.76   & 3.37   & -0.08\\
   0.00  & 5.00&  1.96 &   3.13 &-0.09&  5.00  & 1.87   &3.20        &  -0.06& 5.02  &  1.78    & 3.32    &  -0.08\\
  0.05   & 5.15&   1.99  & 3.10  &  -0.10&  5.15   &  1.88  & 3.19    &  -0.07& 5.17  &  1.80   & 3.28   & -0.08\\
  0.10   & 5.30&  2.00   &3.09    & -0.10&  5.30   &  1.90  & 3.16    &  -0.08 & 5.32  &  1.81   & 3.26    & -0.09\\
  0.20   & 5.60&    2.03  &  3.05 &    -0.11&  5.60   &  1.95 & 3.11   &   -0.10& 5.62  &  1.82 &   3.24     & -0.10
\end{tabular}
\end{ruledtabular}
\end{table*}

In Table \ref{table1} we have gathered the spin magnetic moments
for all cases under study. The total spin moment $M_t$ of the
perfect compounds follows the Slater Pauling behavior being the
number of the valence electrons in the unit cell minus
24.\cite{Galanakis} In the case of the chemically disordered
compounds, doping by 5\%, 10\% or 20\% of Cr (or Fe) atoms, means
that the mean value of the total number of valence electrons in
the unit cell is decreased (or increased respectively) by 0.05,
0.10 and 0.20 electrons, respectively. In most of the cases the
total spin moments follow this behavior a clear sign of the
preservation of the half-metallicity, but in the case of
Co$_2$Mn$_{0.8}$Fe$_{0.2}$Si compound the total moment is 5.14
$\mu_B$ instead of the ideal value of 5.20 $\mu_B$. In the case of
the corresponding Ge and Sn compounds the Fermi level is more deep
in the gap and for the Sn compound it does not cross any more the
minority states. The atom-resolved moments present no peculiarity
and are little sensitive to the doping. Our findings agree with
the conclusions drawn by Kallmayer \textit{et al.} for the
Fe-doped Co$_2$MnSi films.\cite{Kallmayer}

In the second part of our study we study the effect of disorder
between the Mn and the $sp$ atoms. In Fig. \ref{fig2} we present
the atom-resolved DOS for both excess of the $sp$ atom on the left
column and excess of the Mn atoms on the right column. In Table
\ref{table2} we have gathered the total and atomic spin moments
for all cases. Firstly note that the gap is much wider for the Mn
and $sp$ atoms than for the Co atoms since the states around the
gap are of Co-character only. Mixing Mn and $sp$ atoms changes the
symmetry of the Co sites and in this way can induce new states in
the gap and affect the half-metallicity. As shown in Fig.
\ref{fig2}, substituting Si for Mn induces states just at the
right edge of the gap while substituting Mn for Si pushes the
unoccupied minority states even higher in energy and the gap
becomes wider. Overall the DOS is smoothened by the disorder
between the Mn and Si atoms but the main picks do not change
energy position.

In Table \ref{table2} we have gathered the total spin moments for
all cases under study. Substituting 5\%, 10\% or 20\% of the Mn
atoms by the Si, Ge or Sn ones (which are all isoelectronic,
\textit{e.g.} same number of valence electrons) corresponding to
the negative values of $x$ in the table, results in a decrease of
0.15, 0.30 and 0.60 of the total number of valence electrons in
the cell, while the inverse procedure results to a similar
increase of the mean value of the number of valence electrons. The
compounds containing Si and Ge show perfect Slater-Pauling
behavior while the Co$_2$Mn$_{1+x}$Sn$_{1-x}$ deviate from the
ideal values of the total spin moment although in this case the
Fermi level is nearer the center of the gap. Sn is a much heavier
element than both Si and Ge and its mixing with Mn alters
considerably the Coulomb repulsions in the system having a more
profound effect on the half-metallicity of the corresponding
alloy. Thus disorder is more important for the heavy $sp$
elements.

It is interesting also to look at the Mn spin moments.  In the
case of doping presented in the first part of our study doping
scarcely changed the Mn spin moments. Mn atoms remained at the
same sublattice with no immediate change to their close
environment. In the case of disorder excess of  Mn means that Mn
atoms occupy also sites  in the sublattice of the $sp$ atoms while
excess of the $sp$ atoms means that $sp$ atoms are found also in
the sublattice occupied by Mn having  a much larger effect on the
Mn magnetic properties than in the case of doping where Cr and Fe
atoms were found in the Mn-occupied sublattice. As a result the Mn
spin moment can change by as much as $\sim$0.2 $\mu_B$ between the
disordered and the perfect compound.

We have studied the effect of doping and disorder on the magnetic
properties of the Co$_2$MnSi, Co$_2$MnGe, Co$_2$MnSn full-Heusler
alloys. Doping simulated by the substitution of Cr and Fe for Mn
overall keeps the half-metallicity. Its effect depends clearly on
the position of the Fermi level, having the largest one in the
case of Co$_2$MnSi where the Fermi level is near the edge of the
minority-spin gap. On the other hand disorder between the Mn and
the $sp$ atom is more important for the heavy $sp$ atoms like Sn.
Both disorder and doping have little effect on the half-metallic
properties of the compounds which we study and they keep a high
degree of spin-polarization.  It seems that Co$_2$MnGe should be
the most robust compound with respect to its half-metallic
character for experimentalists and realistic applications.


\end{document}